\documentclass[11pt]{article}

\usepackage{latexsym}
\usepackage{mathptmx}
\usepackage[bookmarks]{hyperref}
\usepackage{paralist}
\usepackage{amsfonts}
\usepackage{amsmath}
\usepackage{latexsym}

\usepackage{amsthm}
\usepackage[bookmarks]{hyperref}
\usepackage{comment}

\usepackage[framemethod=default,linewidth=1pt]{mdframed}
\usepackage{appendix}

\usepackage{tikz}
\usetikzlibrary{snakes}
\usetikzlibrary{arrows,shapes}
\definecolor{myblue}{RGB}{20,20,160}
\definecolor{mygreen}{RGB}{20,160,20}
\tikzset{iNode/.style={draw=myblue, rectangle}}
\tikzset{fNode/.style={draw=mygreen, circle}}
\def\nottoobig#1{{\hbox{$\left#1\vcenter to1.111\ht\strutbox{}\right.\n@space$}}}

\newtheorem{theorem}{Theorem}[section]
\newtheorem{corollary}[theorem]{Corollary}
\newtheorem{lemma}[theorem]{Lemma}

\newtheorem{definition}[theorem]{Definition}

\usepackage{epsfig}
\usepackage{amsmath}

\newcommand{\zo}{\{0,1\}}
\newcommand{\prob}{\rm{Prob}}
\newcommand{\mapping}{\rightarrow}
\newcommand{\deb}{\delta\mbox{-BAD}}
\newcommand{\ie}{$\mbox{i.e.}$}

\newcommand{\poly}{\rm poly}

\newcommand{\iid}{$\mbox{i.i.d. }$}
\newcommand{\whp}{$\mbox{w.h.p. }$}

\def\squarebox#1{\hbox to #1{\hfill\vbox to #1{\vfill}}}
\def\qed{\hspace*{\fill}
        \vbox{\hrule\hbox{\vrule\squarebox{.667em}\vrule}\hrule}}

\newenvironment{myproof}{\begin{trivlist}\item[]{\bf Proof}}
                      {\qed \end{trivlist}}

\newenvironment{myprooftwo}{\begin{trivlist}\item[]{\bf Proof}}
                      {\end{trivlist}}

\makeatletter
\def\@listI{\leftmargin\leftmargini \parsep 4.5pt plus 1pt minus 1pt\topsep6pt plus 2pt minus 2pt \itemsep  2pt plus 2pt minus 1pt}

\let\@listi\@listI
\@listi
\makeatother

\author{ {Marius Zimand\/}
\thanks{  Department of Computer and Information Sciences, Towson University,
Baltimore, MD. http://triton.towson.edu/\~{ }mzimand}}

\title{Kolmogorov complexity version of Slepian-Wolf coding}

\date{}
\begin{document}

\maketitle

\begin{abstract}
Alice and Bob are given two correlated $n$-bit strings $x_1$ and, respectively, $x_2$, which they want to losslessly compress and  send to Zack. They can either collaborate by sharing their strings, or work separately. We show that there is no disadvantage in the second scenario: Alice and Bob, without knowing the other party's string,  can compress their strings to almost minimal description length  in the sense of Kolmogorov complexity.  Furthermore, compression takes polynomial time and can be made at any combination of lengths that satisfy some necessary conditions (modulo additive polylogarithmic terms). More precisely, there exist probabilistic algorithms $E_1, E_2$, and deterministic  algorithm $D$, with $E_1$ and $E_2$ running in polynomial time,   having the following behavior:  if  $n_1$, $n_2$  are two  integers satisfying $n_1 + n_2 \geq C(x_1,x_2), n_1 \geq C(x_1 \mid x_2), n_2 \geq C(x_2 \mid x_1)$, then for $i \in \{1,2\}$, $E_i$ on input $x_i$ and $n_i$ outputs a string of length $n_i + O(\log^3 n)$ such that $D$ on input $E_1(x_1), E_2(x_2)$  reconstructs 
$(x_1,x_2)$ with high probability (where $C(x)$ denotes the plain Kolmogorov complexity of $x$, and $C(x \mid y)$ is the complexity of $x$ conditioned by $y$).  Our main result is more general, as it deals with the compression of any constant number of correlated strings. It is an analog in the framework of algorithmic information theory  of the classic Slepian-Wolf Theorem, a fundamental result in network information theory, in which $x_1$ and $x_2$ are realizations of two discrete
random variables representing $n$ independent draws from a joint distribution. In the classical result, the decompressor needs to know the joint distribution of the sources. In our result no type of independence is assumed and the decompressor does not have any prior information about the sources that are compressed.
\end{abstract}

\section{Introduction}

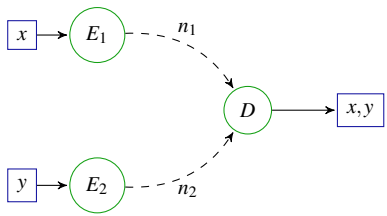
\begin{figure}[h]
\centering
\begin{tikzpicture}[->,>=stealth',shorten >=0.8pt,scale=1.0,auto,node distance=3cm, transform shape]
 \node[iNode] (1) at (1,0) {\scriptsize{$y$}};
  \node[iNode] (2) at (1,2) {\scriptsize{$x$}};
  \node[fNode] (3) at (2, 0) {\scriptsize{$E_2$}};
  \node[fNode](4) at (2,2) {\scriptsize{$E_1$}};
  \node[fNode](5) at (4,1) {\scriptsize{$D$}};
  \node[iNode](6) at (5.5,1) {\scriptsize{$x, y$}};
 \path[every node/.style={font=\sffamily\small}]
    (1) edge node  [left]{}(3)
     (2)    edge node [left]{}(4)
    (3) edge [dashed,bend right] node [below]{\scriptsize{$n_2$}} (5)
    (4) edge [dashed,bend left] node [above]{\scriptsize{$n_1$}} (5)
      (5) edge[] node [left]{} (6);

\end{tikzpicture}
\caption{\scriptsize{Distributed compression: The Slepian-Wolf problem}}
\label{f:figone}
\end{figure}

The Slepian-Wolf Theorem~\cite{sle-wol:j:distribcompression} is the analog of the Shannon's Source Coding theorem for the case of distributed correlated sources. It characterizes the  compression rates for such sources. To illustrate the theorem, let us consider  a data transmission scheme with two senders, Alice and Bob, and one receiver, Zack (see Figure~\ref{f:figone}). Alice has as input an $n$-bit string $x$, Bob has an $n$-bit string $y$. Alice uses the encoding function $E_1 : \zo^n \mapping\zo^{n_1}$ to compress her $n$-bit string to length $n_1$, and sends $E_1(x)$ to Zack. Bob, separately, uses the encoding function $E_2 : \zo^n \mapping\zo^{n_2}$ to compress his $n$-bit string to length $n_2$ and sends $E_2(y)$ to Zack. We assume that the communication channels Alice $\leftrightarrow$ Zack and  
Bob $\leftrightarrow$ Zack are noise-free, and that there is no communication between Alice and Bob. Zack is using a decoding function $D$ and the common  goal of all parties is that $D(E_1(x), E_2(y)) = (x,y)$, for all $x, y$ in the domain of interest (which is defined by the actual model or by the application). In a randomized setting, we allow the previous equality to fail with some small error probability $\epsilon$. Of course, Alice can send the entire $x$ and Bob can send the entire $y$, but this seems to be wasteful if $x$ and $y$ are correlated. We are interested to find what values can $n_1$ and $n_2$ take so that the goal is achieved, when the strings $x$ and $y$ are jointly correlated. 

The Slepian-Wolf theorem takes the standard stance in information theory which assumes that $x$ and $y$ are realizations of some random variables $X$ and, respectively, Y. Furthermore, as it is common in information theory, $(X,Y)$  are assumed to be $2$-Discrete Memoryless Sources ($2$-DMS), which means that $X = (X_1, \ldots, X_n), Y = (Y_1, \ldots, Y_n)$, where the $X_i$'s are \iid Bernoulli random variables, the $Y_i$'s are also \iid Bernoulli random variables, and each $(X_i, Y_i)$ is drawn according to a joint distribution $p(b_1,b_2)$. In other words, 
$(X,Y)$ consists of $n$ independent draws from a joint distributions on pair of bits.
Given the joint distribution $p(b_1,b_2)$ and $X$ and $Y$ of the specified type, the problem amounts to finding the set of values $n_1$ and $n_2$ such that there exists $E_1, E_2$ and $D$ as above with $D(E_1(X), E_2(Y)) = (X,Y)$ with probability converging to $1$ as $n$ grows.  In information theory parlance,  we want to determine the set of achievable transmission rates.   By the Source Coding Theorem, it is not difficult to see that it is necessary that $n_1 \geq H(X \mid Y), n_2 \geq H(Y \mid X)$ and $n_1 + n_2 \geq H(X,Y)$, where $H$ is the Shannon entropy function. The Slepian-Wolf theorem states that these relations are essentially sufficient, in the sense that  any  $(n_1, n_2)$ satisfying strictly the above  three inequalities is a pair of achievable rates, if $n$ is sufficiently large (``strictly" means that ``$>$" replaces ``$\geq$"; see, for example,~\cite{cov-tho:b:inftheory} for the exact statement).

What is surprising is that these optimal achievable rates can be realized with Alice and Bob doing their encoding separately. For example if $H(X)= n, H(Y)=n$, and $H(X,Y) = 1.5n$, then  any pair $(n_1, n_2)$,  with $n_1 > 0.75n, n_2 > 0.75n$,   is a pair of achievable rates, which means that Alice can compress her $n$-bit realization of $X$ to approximately $0.75n$ bits, without knowing Bob's realization of $Y$, and Bob can do the same.  They cannot do better even if they collaborate!

The Slepian-Wolf theorem completely characterizes the set of achievable rates for distributed lossless compression for the case of $2$-DMS, and the result actually holds for an arbitrary number of senders (Theorem~15.4.2,~\cite{cov-tho:b:inftheory}).  However, the type of correlations between $X$ and $Y$  given by the $2$-DMS model is rather simple. In many applications the $(X_i,Y_i)_i$ quantify some  stochastic process at different times $i$ and it is not realistic to assume independence between the values at different $i$'s. The Slepian-Wolf theorem has been extended for sources that are stationary and ergodic ~\cite{cov:j:slepwolfergodic}, but these also capture relatively simple correlations.
\if01
\begin{enumerate}
\item The type of correlations between $X$ and $Y$  given by the $2$-DMS model is rather simple. In many applications the $(X_i,Y_i)_i$ quantify some  stochastic process at different times $i$ and it is not realistic to assume independence between the values at different $i$'s. The Slepian-Wolf theorem has been extended for sources that are stationary and ergodic ~\cite{cov:j:slepwolfergodic}, but these also capture relatively simple correlations.
\item The theorem does not guarantee that the protocol succeeds \whp on all realizations $x$ and $y$ of $X$ and $Y$. In fact, in the current proofs (as far as we know), there are some $(x,y)$, in which the protocol always fails.
\item The encoding and the decoding procedures in the current proofs are inefficient and assume that the senders and the receiver share a public string of exponential length.
\end{enumerate}
The first two issues are inherent to the information-theoretical model.
\fi

 Distributed correlated sources can be alternatively studied using algorithmic information theory, also known as Kolmogorov complexity, which works for individual strings  without any type of independence  assumption, and  in fact without  assuming any generative model that produces the strings. We  recall that $C(u \mid v)$ is the Kolmogorov complexity of $u$ conditioned by $v$, \ie, the length of a shortest program that computes $u$ given $v$ in a fixed universal programming system. $C(u \mid v)$ is also called the \emph{minimum description length} of $u$ given $v$.  If $v$ is the empty string, we simply write $C(u)$ instead of $C(u \mid v)$. One remarkable result in this framework is Muchnik's theorem~\cite{muc:j:condcomp} which states that  there exist algorithms $E$ and $D$ such that for all $n$ and 
for all $n$-bit strings $x$ and $y$,  $E$ on input $x$, $C(x \mid y)$ and $O(\log n)$ help bits outputs a string $p$ of  length $C(x \mid y)$,   and $D$ on input $p$, $y$, and $O(\log n)$ help bits reconstructs $x$. 
Muchnik's theorem relates to the asymmetric version of the above distributed transmission problem in which only Alice compresses her $x$  while Bob  sends the entire $y$  (or, in an equivalent scenario, Zack already knows $y$). 
It says that, given $C(x \mid y)$,  Alice can compute from her string $x$  and only $O(\log n)$ additional help bits 
a string $p$ of minimum description length such that Zack using $p$, $y$ and $O(\log n)$ help bits can reconstruct $x$.
Muchnik's theorem has been strengthened in several ways. Musatov, Romashchenko and Shen~\cite{mus-rom-she:j:muchnik} have obtained a version of Muchnik's theorem for space bounded Kolmogorov complexity, in which both compression and decompression are space-efficient.  Romashchenko~\cite{rom:j:slepwolf} has extended Muchnik's theorem to the general (\ie, non-asymmetric) case. His result is valid for any constant number of senders, but, for simplicity, we present it for the case of two senders: For any two $n$-bit strings $x$ and $y$ and any two numbers $n_1$ and $n_2$ such that $n_1 \geq C(x \mid y)$, $n_2 \geq C(y \mid x)$ and $n_1 + n_2 \geq C(x,y)$, there exist two strings $p_1$ and $p_2$ such that $|p_1| = n_1 + O(\log n), |p_2| = n_2 + O(\log n), C(p_1 \mid x) = O(\log n), C(p_2 \mid y) = O(\log n)$ and $C(x,y \mid p_1, p_2) = O(\log n)$. In words, for any $n_1$ and $n_2$ satisfying the necessary conditions, Alice can compress $x$ to a string $p_1$ of length just slightly larger than $n_1$, and Bob can compress $y$ to a string $p_2$ of length just slightly larger than $n_2$ such that Zack can reconstruct $(x,y)$ from $(p_1, p_2)$, provided all the parties  use a few help bits. These results raise the following questions: (a) can the help bits be eliminated?,
\footnote{In Muchnik's theorem, Alice computes a program $p$ of minimum description length such that $U(p,y)=x$ from $x$, $C(x \mid y)$ and $O(\log n)$ help bits, where $U$ is the universal Turing machine underlying Kolmogorov complexity. One can hope to eliminate the $O(\log n)$ help bits (as we ask in question (a)), but not the $C(x \mid y)$ component. This is not possible even when $y$ is the empty string. Indeed, it is known that for some strings $x$, the computation of $C(x)$ from $x$, and therefore also the computation of a short program $p$ for $x$, requires that some information of size $\log|x|-O(1)$ bits is available~\cite{bau-she:j:compcomp,gac:j:symmetry}.}
and (b) is it possible to  implement the protocol efficiently, 
i.e., in polynomial time?  

Bauwens et al.~\cite{bmvz:c:shortlist}, Teutsch~\cite{teu:j:shortlists} and Zimand~\cite{zim:c:shortlistshortproof}  have obtained versions of Muchnik's theorem with polynomial-time compression, but in which the help bits are still present. In fact, their results are stronger in that the compression  procedure on input $x$ outputs a polynomial-size list of strings guaranteed to contain a short program for $x$ given $y$. This is called list approximation. Note that using $O(\log n)$ help bits, the decoding procedure can pick the right element from the list, re-obtaining Muchnik's theorem. The gain is that this decoding procedure halts even with incorrect help bits, even though the  result may not be the desired $x$.  Next, Bauwens and Zimand~\cite{bau-zim:c:linlist} have eliminated the help bits in Muchnik's theorem, at the cost of introducing a small error probability. Their result can be reformulated as follows.\footnote{In Theorem 3.2 in~\cite{bau-zim:c:linlist}, $y$ is the empty string, but the proof works without modifications for any $y$.} 
\begin{theorem} [\cite{bau-zim:c:linlist}]
\label{t:bz}
There exist a probabilistic algorithm $E$ and a deterministic algorithm $D$ such that $E$ runs in polynomial-time, and for all $n$-bit strings $x$ and $y$ and for every rational number $\delta >0$, 
\begin{enumerate}
\item  $E$ on input $x, 1/\delta$, and $C(x \mid y)$ outputs a string $p$ of length $C(x \mid y) + O(\log^2(n/\delta))$,
\item  $D$ on input $p$  and $y$ outputs $x$, with probability $1-\delta$,
\end{enumerate}
\end{theorem}
Thus in the asymmetric case, Alice can compress her input string in polynomial-time to length which is close to minimum description length (closeness is within a polylog additive term). The decoding algorithm does not run in polynomial time and this is unavoidable if compression is done at this level of optimality because there exist  so called deep strings (these are strings that have short descriptions, but their decompression from short description takes longer than, say, polynomial time).  

In this paper, we prove the analog of Theorem~\ref{t:bz} for the general  non-asymmetric case, \ie, the case in which the number of senders is an arbitrary constant and all senders can  compress their inputs. For simplicity, let us consider again the case with two senders, Alice and Bob, and one receiver, Zack. Alice and Bob are using probabilistic encoding algorithms $E_1$, and respectively $E_2$,  Zack is using the decoding algorithm $D$, and they want that for all $n$, and for all $n$-bit strings $x$ and $y$, $D(E_1(x), E_2(y)) = (x,y)$ with probability $1-\epsilon$. We denote $|E_1(x)|$, the length of $x's$ encoding, and $|E_2(y)|$, the length of $y$'s encoding. How large can these lengths be?  By counting arguments, one can see that
\[
\begin{split}
|E_1(x)|  &\geq C(x \mid y) + \log(1-\epsilon) - O(1) \\
|E_2(y)|  &\geq C(y \mid x) + \log(1-\epsilon) - O(1) \\
|E_1(x)| + |E_2(y)|   &\geq C(x,y) + \log(1-\epsilon) - O(1). \\
\end{split}
\]
Our result implies that the above requirements are also sufficient, except for a small overhead of polylog size. Namely, for any two integers $n_1$ and $n_2$ such that $n_1 \geq C(x \mid y), n_2 \geq C(y \mid x)$ and $n_1 + n_2 \geq C(x,y)$,
 it is possible to achieve $|E_1(x)| \leq n_1 + O(\log^3(n/\epsilon))$, $|E_2(y)| \leq n_2 +  O(\log^3(n/\epsilon))$. Moreover $E_1$ and $E_2$ are polynomial-time probabilistic algorithms. If we do not insist on $E_1$ and $E_2$ running in polynomial time, the overhead can be reduced to
$O(\log(n/\epsilon))$.

For the general case, we need to introduce some notation. Let $\ell$ be the number of senders. For any integers $i$ and $j$, the set $\{1,2, \ldots, i\}$ is denoted $[i]$, and the set $\{i, i+1, \ldots, j\}$ is denoted $[i..j]$ (if $i > j$, this set is empty).  If we have an $i$ tuple of strings $(x_1, \ldots, x_i)$, and $V= \{i_1, i_2, \ldots, i_k\} \subseteq [i]$,  then 
the $k$-tuple $(x_{i_1}, x_{i_2}, \ldots, x_{i_k})$ is denoted $x_V$. 
\begin{theorem}{\bf{(Main Result)}}
\label{t:kolmslepwolf}
There exist probabilistic algorithms $E_1, \ldots E_\ell$, a deterministic algorithm $D$,  and a function $\alpha(n) = \log^{O_\ell(1)} n$ such  that $E_1, \ldots, E_\ell$ run in polynomial time, and 
for every $n$, for every $\ell$-tuple of integers $(n_1, \ldots, n_\ell)$, and for every $\ell$-tuple of $n$-bit strings $(x_1, \ldots, x_\ell)$ if
\begin{equation}
\label{e:constraint}
C(x_V \mid x_{[\ell]-V}) \leq \sum_{i \in V} n_i, \mbox{ for all $V \subseteq [\ell]$},
\end{equation}
then
\begin{itemize}
\item[(a)] For all $i \in [\ell]$, $E_i$ on input $x_i$ and $n_i$ outputs a string $p_i$ of length at most $n_i + \alpha(n)$, 
\item[(b)]  $D$ on input $(p_1, \ldots, p_\ell)$ outputs $(x_1, \ldots, x_\ell)$, with probability $1-1/n$.

\end{itemize}
\end{theorem}
{\bf Notes}
\smallskip

$\bullet$ The constraints~(\ref{e:constraint}) are necessary up to negligible terms.  For example, if there are $\ell=3$ senders, having, respectively, the $n$-bit strings $x_1, x_2$ and $x_3$, and they compress them, respectively,  to lengths
$n_1, n_2$ and $n_3$ and $D(E_1(x_1), E_2(x_2), E_3(x_3)) = (x_1,x_2,x_3)$ with probability $0.99$,  then it is necessary that $n_1 \geq C(x_1 \mid x_2,x_3) - O(1), n_2 \geq C(x_2 \mid x_1, x_3) - O(1),
n_3 \geq C(x_3 \mid x_1,x_2) - O(1), n_1 +n_2 \geq C(x_1,x_2 \mid x_3) - O(1), n_1 +n_3 \geq C(x_1,x_3 \mid x_2) - O(1), n_2 +n_3 \geq C(x_2,x_3 \mid x_1) - O(1)$ and $n_1 +n_2 + n_3 \geq C(x_1,x_2 ,x_3) - O(1)$.
\smallskip


$\bullet$ Compared to Romashchenko's result from~\cite{rom:j:slepwolf},   we have eliminated the help bits, and thus our encoding and decoding is effective.  Moreover, encoding is done in polynomial time (however, as in Theorem~\ref{t:bz} and for the same reason, decoding cannot be done in polynomial time).  The cost is that the encoding procedure  is probabilistic and thus  there is a small error probability. The proof of Theorem~\ref{t:kolmslepwolf} is inspired from Romashchenko's approach, but  the technique is quite different.
\smallskip

$\bullet$ The models in the classical Slepian-Wolf theorem and in Theorem~\ref{t:kolmslepwolf} are different,  and therefore, strictly speaking, the results are not directly comparable. However, there is a relation between Shannon entropy for DMS random variables and the Kolmogorov complexity of the elements in their support. Namely, if $X$ is a DMS, that is it consists of  $n$ independent copies of an $\mbox{i.i.d}$ $\{0,1\}$-valued random variable with distribution $p$, then, for every $\epsilon > 0$, there exists a constant $c_\epsilon$ such that $n H(p) - c_\epsilon \sqrt{n} \leq C(X) \leq n H(p) +  c_\epsilon \sqrt{n}$ with probability $1-\epsilon$. Using this relation, the  classical  theorem can be obtained from the Kolmogorov complexity  version.
\smallskip

$\bullet$ Here are two shortcomings of the classical  Slepian-Wolf Theorem: (a) it assumes strong independence properties of the sources (\ie, the memoryless property), and (b) decompression requires the knowledge of the  distributions of sources. There are versions of this theorem which improve either (a) or (b), but not both. For example, Csisz\'{a}r~\cite{csi:j:universalcoding} has shown  source coding theorems with \emph{universal coding}, which means that the same compression and decompression algorithms work for a large class of sources, without ``knowing" their distributions. But the proof relies on the memoryless property.  Miyake and Kanaya~\cite{miy-kan:j:slepwolfinfspectrum} have obtained a  version of the Slepian-Wolf theorem for general random variables, using   information-spectrum methods  introduced by 
Han and Verd\'{u}~\cite{han-ver:j:infspectrum}. But their proof does not seem to allow universal coding and, moreover, it has an intrinsic asymptotical nature. Theorem~\ref{t:kolmslepwolf} does not require any type of independence, in fact it does not assume any generative model. Also the same compression and decompression algorithms work for all strings satisfying the necessary bounds~(\ref{e:constraint}) \ie, there is universal coding.
\smallskip

 $\bullet$ In the classical Slepian-Wolf theorem, the senders and the receiver share a public string of exponential length.
In Theorem~\ref{t:kolmslepwolf}, the parties do not share any information.
\smallskip

Theorem~\ref{t:kolmslepwolf} is interesting even for the case of a single source compression (i.e., $\ell=1$). 
Note that, by performing an exhaustive search, we obtain a procedure that on input $x$ and $n_1=C(x)$ outputs a shortest program for $x$. However, any such procedure runs in time larger than any computable function~\cite{bau-zim:c:linlist}. In contrast, Bauwens and Zimand (see Theorem~\ref{t:bz}) have shown that if we use randomization, one can find a short program for $x$ in polynomial time, starting with input $(x, n_1= C(x))$. Thus, computing a short program for $x$ from $x$ and $C(x)$ is an interesting example of a task that probabilistically can be done in polynomial time, but deterministically requires time larger than any computable function.  However the requirement that $C(x)$ is known exactly is quite demanding. The following corollary, which is just Theorem~\ref{t:kolmslepwolf} with $\ell=1$, shows that in fact it is sufficient to have an upper bound $n_1 \geq C(x)$. which makes the result more amenable to applications. This solves an open question from~\cite{teu-zim:j:brief}.

\begin{corollary}
There exist a probabilistic algorithm $E$ and a deterministic algorithm $D$ such  that $E$ runs in  polynomial time, and 
for every $n$, for every  $n$-bit string $x$, every positive rational number $\delta>0$, and for every integer $n_1 \geq C(x)$,  
\begin{itemize}
\item[(a)] $E$ on input $x$, $1/\delta$ and $n_1$ outputs a string $p$ of length at most $n_1+ O(\log^3 (n/\delta))$, 
\item[(b)]  $D$ on input $p$ outputs $x$ with probability $1-\delta$.
\end{itemize}
\end{corollary}

\section{Proof of Theorem~\ref{t:kolmslepwolf}}

\subsection{Combinatorial tool: graphs with the rich owner property}
The key tool in the proof is a certain type of bipartite graph, which we call graphs with the rich owner property. Similar graphs, bearing the same name, were used in~\cite{bau-zim:c:linlist}, but the graphs in this paper have a stronger property.
We recall that in a bipartite graph, the nodes are partitioned in two sets, $L$ (the left nodes) and $R$ (the right  nodes), and all edges connect a left node to a right node. We allow multiple edges between two nodes. In all the graphs in this paper, all the left nodes have the same degree, called the left degree.
Specifically, we use bipartite graphs $G$ with $L=\zo^{n}$ , $R=\zo^m$ and with left degree $D=2^d$. We label the edges outgoing from $x \in L$ with strings $z \in \zo^d$. We typically work with a family
of graphs indexed on $n$ and such a family of graphs is
\emph{computable} if there is an algorithm that on input $(x,z)$, where
$x \in L$ and $z \in \zo^d$, outputs the $z$-th neighbor of
$x$. Some of the graphs also depend on a rational $0 < \delta < 1$. A
constructible family of graphs is \emph{explicit} if the above algorithm
runs in time $\poly(n, 1/\delta)$.

We now introduce informally the notions of a \emph{rich owner} and of a \emph{graph with the rich owner property}. Let $B \subseteq L$. The $B$-degree of a right node is the number of its neighbors that are in $B$.  Roughly speaking a left node is a rich owner with respect to $B$,  if most of its right neighbors are ``well-behaved," in the sense that their $B$-degree is not much larger than  $|B| \cdot D /|R|$, the average right degree when the left side is restricted to $B$. One particularly interesting  case, which  is used many times in this paper,   is when most of the neighbors of a left $x$ have $B$-degree $1$, \ie, when $x$ ``owns" most of its right neighbbors. A  graph has the rich owner property if,  for all $B \subseteq L$, most of the left nodes in $B$ are rich owners with respect to $B$. In the formal definition, we replace the average right degree with an arbitrary value, but since in applications, this value is approximately equal to  the average right degree, the  above intuition should be helpful.

The precise definition of a  $(k,\delta)$-rich owner with respect to $B$ is as follows. There are two regimes of interest depending on how large is the size of $B$.   

\begin{definition}  Let $G$ be a bipartite graph as above and let $B$ be a subset of $L$. We say that
$x \in B$ is a $(k,\delta)$-rich owner with respect to $B$ if the following holds:
\begin{itemize}
\item \emph{small regime case:}  If $|B| \leq 2^k$,  then at least $1-\delta$ fraction of $x$'s neighbors  have $B$-degree equal to $1$, that is they are not shared with any other nodes in $B$. We also say that $x \in B$  owns $y$ with respect to B  if $y$ is a neighbor of $x$ and the $B$-degree of $y$ is $1$.
\item  \emph{large regime case:}  If $|B| >  2^k$, then at least a $1-\delta$ fraction of $x$'s neighbors have $B$-degree at most $(2/\delta^2) |B| \cdot D /2^k$.
\end{itemize} 
If $x$
is not a $(k, \delta)$-rich owner with respect to $B$, then it is said to be a $(k,\delta)$-poor owner with respect to $B$.
\end{definition}
\begin{definition}
\label{d:ro}
A bipartite graph $G = (L=\zo^n , R=\zo^m, E \subseteq L \times R)$ has the $(k,\delta)$-rich owner property if 
for every set $B \subseteq L$  all nodes  in $B$, except at most $\delta |B|$ of them,  
 are $(k, \delta)$-rich owners with respect to $B$.
\end{definition}

There are several notions in the literature which are related to our Definition~\ref{d:ro}, the main difference being that they require some non-congestion property similar to rich ownership to hold only for some subsets $B$. Reingold and Raz~\cite{raz-rei:c:extcon} define \emph{extractor-condenser pairs}, in which only subsets  $B$ with size approximately $2^k$ matter.  As already mentioned, Bauwens and Zimand~\cite{bau-zim:c:linlist} use a type of graph also called graphs with the rich owner property, which are close to the extractor-codenser pairs from~\cite{raz-rei:c:extcon}. Capalbo et al.~\cite{cap-rei-vad-wig:c:conductors}  construct \emph{lossless expanders}, which only consider the subsets $B$  in the \emph{small regime case}. In our application, we need to consider subsets $B \subseteq L$ of \emph{any} size and this leads to Definition~\ref{d:ro}, and the distinction between the \emph{small regime case} and the \emph{large regime case}.

The following theorem provides the type of graph that we use. The proof relies on the extractor from~\cite{rareva:c:extractor} and uses a combination of techniques from~\cite{raz-rei:c:extcon},~\cite{cap-rei-vad-wig:c:conductors}, and~\cite{bau-zim:c:linlist}. It is presented  in Section~\ref{s:graph}.
\begin{theorem}
\label{t:richownergraph}
For every natural numbers $n$ and $k$ and for every rational number $\delta \in (0,1]$, there exists an explicit  bipartite graph $G = (L,R, E \subseteq L \times R)$ that has the $(k,\delta)$-rich property with the following parameters:

\begin{enumerate}[\scshape (i)]
\item $L = \zo^n$,
\item  $R = \zo^{k+ \gamma(n/\delta)}$,
\item left degree $D = 2^{\gamma(n/\delta)}$, 
\end{enumerate}
where $\gamma(n) = O(\log^3(n/\delta))$.

\end{theorem}

\subsection{Proof overview}
\label{s:overview}
For this proof sketch, we consider the case with $\ell=2$ senders, which have as input the $n$-bit strings $x_1$ and, respectively, $x_2$. By hypothesis, the compression lengths $n_1$ and $n_2$ satisfy
\[
n_1 \geq C(x_1 \mid x_2), n_2 \geq C(x_2 \mid x_1), n_1 + n_2 \geq C(x_1, x_2).
\]
 The two senders use graphs $G_1$ and, respectively,  $G_2$,   with the $(n_1 + 1, \delta)$ and, respectively, $(n_2 + 1, \delta)$-rich owner property and with $\delta = 1/n^2$,  obtained from Theorem~\ref{t:richownergraph}. The left nodes in both graphs are the set of $n$-bit strings, the right nodes in $G_1$ are the binary strings of length $n_1 + \gamma(n/\delta)$, and  the right nodes in $G_2$ are the binary strings of length $n_2 + \gamma(n/\delta)$.  Sender $1$ picks $p_1$, a random neighbor of $x_1$ (viewed as a left node) in $G_1$, and  sender $2$ picks $p_2$, a random neighbor of $x_2$ (viewed as a left node) in $G_2$.  
 
 We need to explain how the receiver can reconstruct $x_1$ and $x_2$ from $p_1$ and $p_2$. Most of the statements below hold with probability $1-O(\delta)$. For conciseness, when this is clear, we omit mentioning this fact. We first assume that the decompression procedure knows $C(x_1), C(x_2)$ and $C(x_1, x_2)$ (this is usually called the complexity profile of $x_1$ and $x_2$). We will see later how to eliminate this assumption.
 
 The first case to analyze is when $C(x_2) \leq n_2$.  Then $x_2$ can be constructed as follows. Let  $B = \{x \in \zo^n \mid C(x) \leq C(x_2)\}$. This is a subset of the left nodes in $G_2$, that contains $x_2$, and is in the \emph{small regime case} (because $|B| <  2^{C(x_2)+1} \leq 2^{n_2+1}$). The set of poor owners in $G_2$ w.r.t. $B$  has size at most $\delta\cdot |B| = 2^{C(x_2)- \log(1/\delta)}$. Since the set of poor owners w.r.t. $B$ can be effectively enumerated given $C(x_2)$, we derive that every  poor owner has complexity less than $C(x_2)$. So,  $x_2$ is a rich owner, which implies that with probability $1-\delta$, $x_2$ does not share $p_2$ with any other nodes in $B$. It follows that $x_2$ can be constructed from $p_2$ by enumerating $B$ till we encounter a neighbor of $p_2$. As we have seen,  with probability $1-\delta$, this neighbor is $x_2$. Next, we take $B = \{x_1' \in \zo^n \mid C(x_1' \mid x_2) \leq C(x_1 \mid x_2)\}$, and in a similar way we show that $B$ is in the \emph{small regime case} in $G_1$, and $x_1$ is a rich owner w.r.t. $B$. Therefore, with probability $1-\delta$,  $x_1$  owns $p_1$.  Thus, if we enumerate $B$ till we encounter a neighbor of $p_1$, we obtain $x_1$.  

The other case is when $C(x_2) > n_2$. We can show that with high probability, 
\begin{equation}
\label{e:over1}
C(p_2) =^*  n_2,
\end{equation}
 where $=^*$ means that the equality holds up to poly-logarithmic terms; we use $\leq^*$ and $\geq^*$ in a similar way.  For that, again we consider $B = \{x \in \zo^n \mid C(x) \leq C(x_2)\}$. This is a subset of the left nodes of $G_2$ that is now in the \emph{large regime case}. In the same way as above,  $x_2$ is a rich owner in $G_2$ w.r.t. $B$, which implies that with probability $1-\delta$, it shares $p_2$ with at most $(2/\delta^2) |B|D/2^{n_2} = 2^{C(x_2)- n_2 +{\rm poly}(\log n)}$ other nodes in $B$. Taking into account that $B$ can be enumerated given $C(x_2)$, it follows that $x_2$ can be constructed from $p_2$, $C(x_2)$, and $x_2$'s rank among $p_2$'s neighbors in $B$, which implies that
$C(x_2) \leq^* C(p_2) + (C(x_2) - n_2)$. So, $C(p_2) \geq^{*} n_2$. Since the length of $p_2$ is $=^* n_2$, we derive that $C(p_2) =^* n_2$.

The next observation is that, given $p_2, x_1$ and $C(x_2 \mid x_1)$, the receiver can construct $x_2$. At this moment, the receiver does not have $x_1$, so actually $x_2$ will be constructed later, after the receiver has $x_1$.  
However, the observation is helpful even at this stage. Let us first see why the observation is true. Consider $B = \{x_2'  \in \zo^n \mid C(x_2' \mid x_1) \leq C(x_2 \mid x_1)\}$. This is a subset of left nodes in $G_2$ that contains $x_2$, and is in the \emph{small regime case} (because $|B| <  2^{C(x_2\mid x_1)+1} \leq 2^{n_2+1}$). Similarly to the argument used earlier, $x_2$ is a rich owner w.r.t. $B$. So, $x_2$ owns $p_2$ w.r.t. $B$, which implies that $x_2$ can be obtained by enumerating the elements of $B$ till we encounter one that is a neighbor of $p_2$.

The observation implies that $C(x_2, x_1) \leq^* C(p_2, x_1)$. Since it also holds that $C(p_2, x_1) \leq^* C(x_2, x_1)$ (because $p_2$ can be obtained from $x_2$ and its index among $x_2$'s neighbors in $G_2$, which takes poly log $n$ bits to describe), we have 
\begin{equation}
\label{e:over2}
C(x_2, x_1) =^* C(p_2, x_1).
\end{equation}
Then,  by~(\ref{e:over1}) and (\ref{e:over2}),  
\begin{equation}
\label{e:over3}
C(x_1 \mid p_2) =^* C(x_1, p_2) - C(p_2) =^* C(x_1, x_2) - n_2,
\end{equation}
where the first $=^*$ follows from the chain rule. The last estimation, allows the receiver to reconstruct $x_1$ from $p_1$ and $p_2$. For that, consider $B = \{x_1' \in \zo^n \mid C(x_1' \mid p_2) \leq C(x_1, x_2) - n_2 \}$.  Our estimation~(\ref{e:over3}) of $C(x_1 \mid p_2)$ implies that $x_1$ is in $B$ (in this proof sketch we ignore the * in equation~(\ref{e:over3})). Next, by the same argument as above, the poor owners in $G_1$  have complexity conditioned by $p_2$ less than $C(x_1, x_2) - n_2$, and this implies that $x_1$ is not a poor owner. Since $C(x_1, x_2) - n_2 \leq (n_1+n_2) - n_2 = n_1$, $B$ is in the \emph{small regime case}. This implies that with high probability $x_1$ owns $p_1$ in $G_1$ w.r.t. $B$. So, if we enumerate $B$ till we encounter a neighbor of $p_1$, we obtain $x_1$.

With $x_1$ in hand, the receiver constructs $x_2$, using  the earlier observation.
\smallskip

{\bf Decompression without knowledge of the input's complexity profile.}  As promised, we show how to eliminate the assumption that the decompressor $D$ knows $C(x_1), C(x_2), C(x_1,x_2)$.  The idea is to let  $D$ run the above procedure for all possibilities of $C(x_1), C(x_2), C(x_1,x_2)$ and use hashing to isolate the correct run (or some run that produces the same output). Since $x_1$ and $x_2$ are $n$-bit strings, there are $O(n^3)$ possibilities for the triplet $(C(x_1), C(x_2), C(x_1,x_2))$ and hashing will add only $O(\log n)$ bits. For hashing we use the following result. Alternatively, it is possible to use the almost $\delta$-universal function of Naor and Naor~\cite{na-na:j:smallbias}, or Krawczyk~\cite{kra:c:hash}.
\begin{lemma}[~\cite{bau-zim:c:linlist}]
\label{l:mod}
Let $x_1, x_2 \ldots, x_s$ be distinct $n$-bit strings, which we view in some canonical way as integers $< 2^{n+1}$.

Let $q_i$ be the $i$-th prime number and let $L = \{q_1, \ldots, q_t\}$, where $t = (1/\delta) \cdot s \cdot n$.

For every $i \leq s$, for $(1-\delta)$ fraction of $q$ in $L$, the value of $x_i \hspace{-0.2cm}\mod q$ is unique in the sequence 
$(x_1 \bmod q, x_2 \bmod q, \ldots,  x_s \bmod q)$.
\end{lemma}
For $i=1,2$, Sender $i$ who has input $x_i$ will send in addition to $p_i$ (a random neighbor of $x_i$ in $G_i$, as we have seen above), also  the string hash$(x_i)$, which is computed as follows. Taking into account that for any $n$-bit string $u$, $C(u) \leq |u|+O(1)$, we
let $s = O(n^3)$ be an upper bound  for the number of all triplets $(C(u), C(v), C(u,v))$, where $u$ and $v$ are $n$-bit strings,  and let  $t= (1/\delta) \cdot s \cdot n$. Now, hash$(x_i) =
(q_i, x_i \bmod q_i)$,  where $q_i$ is a prime number chosen at random from the first $t$ prime numbers, 
The decompressor runs \emph{in parallel} the procedure presented above for all $s$ guesses for $(C(x_1), C(x_2), C(x_1,x_2))$ and halts when the first of the parallel runs outputs $x'_1, x'_2$ with $x'_1 \bmod q_1 = x_1 \bmod q_1$ and $x'_2 \bmod q_2 = x_2 \bmod q_2$.  Note that some of the parallel runs may not halt, but the run corresponding to the correct guess of $(C(x_1), C(x_2), C(x_1,x_2))$  halts and yields, as we have seen, $(x_1, x_2)$ with probability $1-O(\delta)$. By Lemma~\ref{l:mod}, the probability that a run halts with $x'_1 \not= x_1$ or $x'_2 \not= x_2$ but $x'_1 \bmod q_1 = x_1 \bmod q_1$ and $x'_2 \bmod q_2 = x_2 \bmod q_2$ is at most $\delta$. Consequently, this procedure reconstructs correcty $(x_1, x_2)$ with probability $1-O(\delta)$. Since the $t$-th prime number is bounded by $t \log t$ and can be found in time polynomial in $t$,  the length of each of the compressed strings increases with only  $O(\log t) = O(\log (n/\delta))$ bits, and the running time of compression is still polynomial. 
\smallskip

If the number of senders is $\ell > 2$, several technical complications arise. In the case $\ell=2$, sketched above, the decoding algorithm needs to have $C(x_1), C(x_2)$ and $C(x_1, x_2)$ to be able to enumerate the various sets $B$. As we have seen, we can assume that the receiver knows the complexity profile of the input strings, and therefore, the decoding algorithms has these values. When $\ell \geq 3$, the various sets $B$ are defined in term of complexities containing certain combinations of the input strings $x_i'$s, and of the randomly picked right neighbors, $p_j$'s. To give just one example, the complexity $C(x_{[k]}, p_{[k+1..\ell]})$ is required at some point. The decoding algorithm needs to obtain, with high probability, good approximations of such complexities from the complexity profile of the input strings (see Lemma~\ref{l:alg}). Another technical aspect is that the approximation slacks (hidden above in the notation $=^*, \leq^*, \geq^*$, and also those arising in the estimations of the complexities of ``combined" tuples of $x_i$'s and $p_j$'s) cannot be ignored as we did in this proof sketch. To handle this, senders use graphs with decreasing $\delta$'s  (\ie, $\delta_\ell > \delta_{\ell-1} > \ldots > \delta_1$) and increasing overhead in the length of the right neighbors. More precisely, sender $k$ (for every $k \in [\ell]$),  uses a graph $G_k$ with the $\delta_k$-rich neighbor property, in which the right nodes have length $n_k + \gamma(n/\delta_k) + \eta_k(n)$, where the additional $\eta_k(n)$ is needed to handle the effect of approximations.  The overhead $\gamma(n/\delta_k) + \eta_k(n)$ is bounded by $(\log n)^{O_\ell(1)}$, where $O_\ell(1)$ denotes a constant that depends on $\ell$.  In spite of these technicalities, the core ideas of the proof  are those presented in the above sketch.

\subsection{Parameters}
\smallskip

We fix $n$, the length of the input strings $x_1, \ldots, x_\ell$.

We use a constant $c$ that will take  a large enough value so that the estimations done in this proof  are all valid. The construction uses parameters $(\delta_\ell, \delta_{\ell-1}, \ldots, \delta_1)$ , $(\gamma_\ell, \gamma_{\ell-1}, \ldots, \gamma_1)$,  $(\eta_\ell, \eta_{\ell-1}, \ldots, \eta_1)$ and  $(\hat{\delta}_\ell, \hat{\delta}_{\ell-1}, \ldots, \hat{\delta}_1)$ that are all functions of $n$ and are defined as follows. 
\smallskip

$\bullet$   For all $k \in [\ell]$, $\gamma_k$ is defined in terms of $\delta_k$, according to the relation given in Theorem~\ref{t:richownergraph}: $\gamma_k = O( \log^3 (n/\delta_k))$. We also define $\gamma_{\ell+1}=0$. 
\smallskip

$\bullet$ The parameters $\delta_k$ are defined recursively in descending order as follows: $1/\delta_\ell = c \cdot n$, and then $1/\delta_k = 2^{17\gamma_{k+1}}$, for $k =\ell-1, \ldots, 1$.  
Note that for  all $k \in [\ell - 1]$,  $(1/\delta_k) = 2^{ (\log n)^{O_\ell(1)}}$, $\gamma_k = O(\log^3 n \cdot \gamma^3_{k+1})$  and $\gamma_k = (\log n)^{O_\ell(1)}$, where $O_\ell(1)$ denotes a constant that depends on $\ell$.
We will use the fact that for any constant $a$, the following inequalities hold provided $n$ is large enough:
\begin{equation}
\label{e:gamma}
\gamma_k  \geq a (\gamma_{k+1}  + \log(1/\delta^2_k) +  \log n),
\end{equation} 
and
\begin{equation}
\label{e:delta}
\log(1/\delta_k) > 16\gamma_{k+1} + a \log n.  
\end{equation} 
\smallskip

$\bullet$ We next define for all $k \in [\ell]$,
\begin{equation}
\label{e:eta}
\eta_k = 2 \gamma_k + \log(2/\delta^2_k).
\end{equation}
Note that for all $k$,  $\eta_k =  (\log n)^{O_\ell(1)}$.
\smallskip

$\bullet$ We denote $\hat{n}_k = n_k + \eta_k +1$.
\smallskip

$\bullet$  The sequence $\hat{\delta}_\ell, \hat{\delta}_{\ell-1}, \ldots, \hat{\delta}_1$ is defined recursively (in descending order) as follows: $\hat{\delta}_\ell = \delta_\ell$ and
\[
\hat{\delta}_k = 2 \hat{\delta}_{k+1} + \delta_{k}.
\]
It can be checked that $(1/\hat{\delta}_k) =  2^{ (\log n)^{O_\ell(1)}}$
\smallskip
\subsection{Handling the input complexity profile}

As we did in Section~\ref{s:overview}, Proof overview, we first assume that the decompressor $D$ knows the complexity profile of the input strings $x_1, \ldots, x_\ell$, which is the tuple $(C(x_V) \mid V \subseteq [\ell], V \not= \emptyset)$. This assumption can be eliminated in the same way as we did in the proof overview.
 
\subsection{Encoding}


Each sender $k$, $k \in [\ell]$, has as input the $n$-bit string $x_k$ and uses the graph $G_k$ promised by Theorem~\ref{t:richownergraph}, with $L = \zo^n, R= \zo^{\hat{n}_k+\gamma_k}$ that has the $(\hat{n}_k, \delta_k)$-rich owner property. Thus left nodes are $n$-bit strings and in this way the input string $x_k$  is a left node  in $G_k$.   Sender $k$ picks $p_k$  uniformly at random among the right neighbors of $x_k$ in the graph $G_k$, and sends $p_k$ to the receiver. The length of $p_k$ is $\hat{n}_k + \gamma_k = n_k + (\log n)^{O_\ell(1)}$.  (If the length $n$ of the input strings is not known by the receiver, Sender $k$ also sends the length of $x_k$. Note that the algorithms work even if the strings $x_1, \ldots x_\ell$ have different lengths, in which case in the proof $n$ is the maximum of these lengths.)

\subsection{Decoding} 
We first state some technical lemmas that play an important role in the decoding procedure.  They are proved in Section~\ref{s:technicallemmas}.
The first two lemmas estimate how the complexity of $p_k$ is related to the complexity of $x_k$, for $k \in [\ell]$. There are two regimes to analyze, depending on whether the complexity of  $x_k$ is low or high. We analyze the respective complexities conditioned by some string $b$, which for now is an arbitrary string, but later when we apply these lemmas for $p_k$ and $x_k$, $b$ will be 
instantiated with the previous inputs $x_{[k-1]}$ and the nodes $p_{[k+1..\ell]}$.
\begin{lemma} 
\label{l:small}
\emph{(low complexity case)}  Let $b$ be an arbitrary string and suppose $C(x_k \mid b) \leq n_k + \eta_k$.
\begin{enumerate}[\scshape (i)]
\item There exists an algorithm that on input $b, p_k$ and $C(x_k \mid b)$ outputs $x_k$ with probability $1-\delta_k$ (over the random choice of $p_k$).
\item With probability $1-\delta_k$, $\big| C(p_k, b) - C(x_k, b) \big| \leq \gamma_k  + O(\log n) =  (\log n)^{O_\ell(1)}$.
\end{enumerate}
\end{lemma}

\begin{lemma}
\label{l:large}
\emph{(high complexity case)} Let $b$ be an arbitrary string and suppose $C(x_k \mid b) >  n_k $.
\begin{enumerate}[\scshape (i)]
\item There exists an algorithm that on input $b, p_k$, $C(x_k \mid b)$  and  some string $b'$ of length 
$|b'| \leq \max(0, C(x_k \mid b) - (n_k + \eta_k - \gamma_k - \log(2/\delta^2_k)))$, outputs $x_k$ with probability $1-\delta_k$
(over the random choice of $p_k$).
\item With probability $1-\delta_k$, $\big| C(p_k,b)  -  (C(b) + n_k + \eta_k) \big| \leq \gamma_k + \eta_k + \log(2/\delta^2_k)+ O(\log n) =  (\log n)^{O_\ell(1)}$ and
 $C(p_k \mid b) \geq n_k + \eta_k - \gamma_k - \log(2/\delta^2_k) - O(\log n) = n_k -  (\log n)^{O_\ell(1)}$.
\end{enumerate}
\end{lemma}

The decoding procedure needs good estimations of the complexities of the form $C(x_k \mid x_{[k-1]}, p_{[k+1..\ell]})$. The following lemma shows that it is possible to effectively approximate them with precision $(\log n)^{O_\ell(1)}$.
The inductive proof  requires the approximation of more general complexities
of the form $C(x_V, p_{[k+1..\ell]})$ for all $V \in {\cal P}([k])$ and for all $k \in [\ell]$.
\begin{lemma}
\label{l:alg}
There is an algorithm with the following behaviour.


For all $k \leq \ell$, the algorithm on input   $(p_{k+1}, \ldots, p_\ell)$, $V \in {\cal P}([k])$,  and $C(x_W)$ for all non-empty $W \subseteq [\ell]$,   outputs an integer $A(x_V, p_{k+1}, \ldots, p_\ell)$ such that with probability
$1-\hat{\delta}_k$,
\[
\big| C(x_V, p_{[k+1..\ell]}) - A(x_V, p_{[k+1..\ell]}) \big|  \leq 4 \gamma_{k+1} =  (\log n)^{O_\ell(1)}.
\]
\end{lemma}

The next lemma shows that the constraints~(\ref{e:constraint})  remain roughly valid if we replace the left nodes $x_{k+1}, \ldots, x_\ell$ with the corresponding right nodes $p_{k+1}, \ldots, p_\ell$.
\begin{lemma}
\label{l:bound}
For all $k \leq \ell$, for all non-empty $V \subseteq [k]$,  the following inequality holds with probability $1- \hat{\delta}_k$:
\[C(x_V \mid x_{[k]-V}, p_{[k+1..\ell]}) \leq \sum_{j \in V} n_j + O(\ell-k) \cdot \log n
\]
\end{lemma}
\bigskip

{\bf Decoding algorithm}
\bigskip

Some of the estimations below hold with error probability  bounded by $\hat{\delta}_k$ or $\delta_k$, for various $k \in [\ell]$, and all these values are bounded by  $\delta_\ell = 1 /(c \cdot n)$  (the probability is on the random choices of $p_1, \ldots, p_\ell$). There are $O_\ell(1)$  ``bad" events when the estimations are violated. By taking $c$ sufficiently large, the union of all ``bad events"  has probability at most $1/n$. The following arguments are done conditioned on the event that none of the  ``bad" events happened.

First, using the algorithm from Lemma~\ref{l:alg},  the values $A(x_k \mid x_{[k-1]}, p_{[k+1..\ell]})$ are calculated by the formula
\[
A(x_k \mid x_{[k-1]}, p_{[k+1..\ell}]) = A(x_k,  x_{[k-1]}, p_{[k+1..\ell]}) - A(x_{[k-1]}, p_{[k+1..\ell]}).
\]
By the chain rule and the bounds on approximation error established  in Lemma~\ref{l:alg}, it holds that
\begin{equation}
\label{e:ck}
\big| C(x_k \mid x_{[k-1]}, p_{[k+1..\ell]}) - A(x_k \mid x_{[k-1]}, p_{[k+1..\ell]}) \big| \leq 8\gamma_{k+1} + O(\log n).
\end{equation}
 The decoding algorithm reconstructs in order $x_1, x_2, \ldots, x_\ell$.
\smallskip

{\bf Step 1 }(reconstruction of $x_1$).
\smallskip

By Lemma~\ref{l:bound}, 
\begin{equation}
\label{e:cxn}
C(x_1 \mid p_{[2..\ell]}) \leq n_1 + O(\ell-2) \log n.
\end{equation}
Consider the graph $G_1 = (L,R, E\subseteq L \times R)$ used by sender $1$. $G_1$ has $L=\zo^n, R = \zo^{\hat{n}_1 + \gamma_1}$, left degree $D= 2^{\gamma_1}$, and the $(\hat{n}_1, \delta_1)$-rich owner property.
Consider the set
\[
B = \{x \in \zo^n \mid C(x \mid p_{[2..\ell]}) \leq A(x_1 \mid p_{[2..\ell]}) + 8 \gamma_2 + O(\log n)\},
\]
where the constant hidden in the $O( )$ is taken so that $x_1$ is in $B$ (keeping in mind the estimation~(\ref{e:ck}) for $k=1$).

The subset of $(\hat{n}_1, \delta_1)$-poor owners w.r.t. $B$ in $G_1$ has size at most $\delta_1 \cdot |B| \leq \delta_1 \cdot 2^{A(x_1 \mid p_{[2..\ell]} )+ 8 \gamma_2 + O(\log n))}$. Note that the set of poor owners can be enumerated given
$n$, $p_{[2..\ell]}$, $A(x_1 \mid p_{[2..\ell]})$, $\hat{n}_1$, and $\delta_1$. Given $p_{[2..\ell]}$,  $A(x_1 \mid p_{[2..\ell]})$ can be computed from $n$ and the complexity profile of the input strings (by Lemma~\ref{l:alg}).  The integers  $\hat{n}_1$ and $\delta_1$ can be computed from $n$ and $n_1$ and we can assume that $n_1 \leq n$ (otherwise, sender $1$ 
can simply send $x_1$ uncompressed).  It follows that if $x$ is a poor owner, then
\[
\begin{split}
C(x  \mid p_{[2..\ell]}) & \stackrel{(a)}{\leq} \log(\delta_1 \cdot |B|) + O(\log n) \\
&\stackrel{(b)}{\leq} A(x_1 \mid p_{[2..\ell]})+ 8 \gamma_2 - \log(1/\delta_1) + O(\log n) \\
&  \stackrel{(c)}{\leq} C(x_1 \mid p_{[2..\ell]}) + 16 \gamma_2 - \log(1/\delta_1) + O(\log n) \\
& \stackrel{(d)}{<} C(x_1 \mid p_{[2..\ell]}).
\end{split}
\]
Transition (a) follows taking into account the above explanations and the fact that $x$ is described by its index in the enumeration of poor owners, transition (b) uses the above bound for the number of poor owners, 
transition (c) follows from~(\ref{e:ck}), and transition (d) follows from~(\ref{e:delta}).

Therefore, $x_1$ cannot be a poor owner, so it is a $(\hat{n}_1,\delta_1)$-rich owner in $G_1$. The size of $B$ is bounded by $2^{n_1+\eta_1+1}$ because
\[
\begin{split}
A(x_1 \mid p_{[2..\ell]}) +8 \gamma_2 + O(\log n) & \stackrel{(a)}{\leq} C(x_1 \mid p_{[2..\ell]}) + 16 \gamma_2 + O(\log n) \\
& \stackrel{(b)}{\leq} n_1 + 16 \gamma_2 + O(\ell-2) \log n \\
&\stackrel{(c)}{<} n_1 + \eta_1 +1 = \hat{n}_1.
\end{split}
\]
Transition (a) follows from~(\ref{e:ck}), transition (b) follows from~(\ref{e:cxn}), and transition (c) follows from~(\ref{e:eta}) and~(\ref{e:gamma}).

Hence $B$ is  in the  \emph{small regime case} for the graph $G_1$. It follows that with probability $1-\delta_1$, $x_1$ is the only node in $B$ that is a neighbor of $p_1$ in $G_1$. Therefore, $x_1$ can be reconstructed as follows: Enumerate
$B$ till we encounter one element that is a left neighbor of $p_1$ in $G_1$ and output this element. By the above discussion, this procedure will output $x_1$ with high probability.
\smallskip

{\bf Step k} (we have already obtained $x_1, \ldots, x_{k-1}$ and now we reconstruct $x_k$).
\smallskip

The argument is similar to the one in Step 1. By Lemma~\ref{l:bound}, $C(x_k \mid x_{[k-1]}, p_{[k+1..\ell]}) \leq n_k + O(\ell-k) \log n$.
Consider the graph $G_k = (L,R, E\subseteq L \times R)$ used by sender $k$. $G_k$ has $L=\zo^n, R = \zo^{\hat{n}_k + \gamma_k}$, left degree $D= 2^{\gamma_k}$, and the $(\hat{n}_k, \delta_k)$-rich owner property.
Consider the set
\[
B = \{x \in \zo^n \mid C(x \mid x_{[k-1]}, p_{[k+1..\ell]}) \leq A(x_k \mid x_{[k-1]},  p_{[k+1..\ell]}) + 8 \gamma_{k+1} + O(\log n)\},
\]
where the constant hidden in the $O( )$ is taken so that $x_k$ is in $B$ (keeping in mind the estimation~(\ref{e:ck}) ). Using a similar argument as in Step $1$, $x_k$ is a $(\hat{n}_k, \delta_k)$-rich owner w.r.t $B$ in $G_k$ and $B$ is in the  \emph{small regime  case}, because
\[
\begin{split}
A(x_x \mid x_{[k-1], } p_{[k+1..\ell]}) + 8 \gamma_{k+1} + O(\log n) & \stackrel{(a)}{\leq} C(x_k \mid x_{[k-1]}, p_{[k+1..\ell]}) + 16 \gamma_{k+1}+ O(\log n) \\
& \stackrel{(b)}\leq n_k + 16 \gamma_{k+1} + O(\ell-k) \log n \\
& \stackrel{(c)} < n_k + \eta_k + 1 = \hat{n}_k.
\end{split}
\]
Transition (a) follows from~(\ref{e:ck}), transition (b) follows from~Lemma~\ref{l:bound}, and transition (c) follows from~(\ref{e:eta}) and~(\ref{e:gamma}).
 Therefore, similarly to Step $1$, $x_k$ can be obtained from $x_{[k-1]}, p_k$, and $p_{[k+1..\ell]}$, because $x_k$ owns $p_k$ w.r.t.  $B$ in $G_k$, and $B$ can be enumerated given $x_{[k-1]}$, and $p_{[k+1..\ell]}$.

\section{Proofs of the technical lemmas}
\label{s:technicallemmas}

This section contains the proofs of Lemma~\ref{l:small}, Lemma~\ref{l:large}, Lemma~\ref{l:alg}, and Lemma~\ref{l:bound}.
\smallskip

\begin{myproof} {\bf of Lemma~\ref{l:small}.}
(i) The graph $G_k = (L_k,R _k , E_k \subseteq L_k \times R_k)$, used by sender $k$ for doing the encoding is obtained by applying Theorem~\ref{t:richownergraph} with parameters $n, k = n_k + \eta_k + 1 =  \hat{n}_k$ 
and $\delta_k$, and thus has $L_k = \zo^n, R_k= \zo^{\hat{n}_k + \gamma_k}$ and the $(\hat{n}_k, \delta_k)$-rich owner property.
Let
\[
B = \{x \in \zo^n \mid C(x \mid b) \leq C(x_k \mid b)\}.
\]
Note that $B$'s size is bounded by $2^{C(x_k \mid b)+1}$ and, obviously, $x_k \in B$.

The subset of poor owners w.r.t. $B$ has size at most $\delta |B| \leq \delta_k \cdot  2^{C(x_k \mid b)+1}$ and can be enumerated given $b$ and $C(x_k \mid b)$. It follows that if $x$ is a poor owner w.r.t. $B$, then
\[
\begin{array}{ll}
C(x \mid b) &\leq C(x_k \mid b)+ 1 -  \log(1/\delta_k) + O(\log n) \\
                   & < C(x_k \mid b),
\end{array}
\]
where the second inequality  holds because $1/\delta_k \geq 1/\delta_\ell = c n $ and $c$ is chosen to be a large enough constant. Consequently,  $x_k$  cannot be a poor owner, and therefore it is a $(\hat{n}_k, \delta_k)$-rich owner w.r.t. $B$. 
 Since $|B| \leq 2^{C(x_k \mid b) +1}$ and $C(x_k \mid b) + 1 \leq n_k + \eta_k +1 = \hat{n}_k$, we are in the  \emph{small regime case}.  By the property of graphs with the rich owner property in this regime of parameters, it follows that with probability $(1-\delta_k)$, $x_k$ is the only node in $B$ that is a neighbor of $p_k$. This leads to the following algorithm that constructs $x_k$, on input $b, p_k$ and $C(x_k \mid b)$: Enumerate $B$ till one of the enumerated nodes is a neighbor of $p_k$. As we have seen, with probability $1-\delta_k$, this node is $x_k$.
 
 (ii)  It follows from (i) that, with probability $1-\delta_k$,
 \[
 \begin{array}{ll}
 C(x_k , b) & \leq C(p_k, b) + 2 \log C(x_k \mid b )+ O(1) \\
 & \leq C(p_k, b) + 2 \log n + O(1).
 \end{array}
 \]
Since $C(p_k, b) \leq C(x_k, b) +  \gamma_k +O(\log n)$ (because $p_k$ can be obtained from $x_k$ and the index of the edge that links $x_k$ and $p_k$ among the edges going out from $x_k$; next, we take into account that the left degree of $G$ is $2^{\gamma_k}$ and consequently the index requires $\gamma_k$ bits), the conclusion follows.
\end{myproof}
\if01
\begin{lemma}
\label{l:large}
\emph{(Large regime)} Let $b$ be an arbitrary string and suppose $C(x_k \mid b) >  n_k $.
\begin{enumerate}[\scshape (i)]
\item There exists an algorithm that on input $b, p_k$, $C(x_k \mid b)$  and  string $b'$ of length $|b'| \leq C(x_k \mid b) - (n_k + \eta_k)$, outputs $x_k$ with probability $1-\delta_k$
(over the random choice of $p_k$).
\item With probability $1-\delta_k$, $\big| C(p_k,b)  -  (C(b) + n_k + \eta_k) \big| \leq \gamma_k + O(\log n)$ and $C(p_k \mid b) \geq n_k + \eta_k - O(\log n)$.
\end{enumerate}
\end{lemma}
\fi
\begin{myproof} {\bf of Lemma~\ref{l:large}.} There are two cases to analyze: \emph{Case 1:} $C(x_k \mid b) \in (n_k, n_k + \eta_k]$ and 
\emph{Case 2:} $C(x_k \mid b) > n_k + \eta_k$.

In \emph{Case 1}, the same estimations as in  Lemma~\ref{l:small} hold, because we are still in the \emph{small regime case.}  Thus, we obtain
\[
C(x_k \mid b) - 2 \log n - O(1) \leq C(p_k \mid b) \leq C(x_k \mid b) + \gamma_k + O(\log n).
\]
Using the fact that we are in Case 1, we can substitute $C(x_k \mid b)$ and obtain
\[
n_k + \eta_k - (\eta_k + O(\log n)) \leq C(p_k \mid b) < n_k + \eta_k + (\gamma_k + O(\log n)).
\] 
Using the chain rule, we  obtain
\[
C(b) + n_k + \eta_k - (\eta_k + O(\log n)) \leq C(p_k,  b) < C(b) + n_k + \eta_k + (\gamma_k + O(\log n)),
\]
which implies
\begin{equation}
\label{e:eqcaseone}
\big| C(p_k, b)  - (C(b) + n_k + \eta_k)\big| \leq \eta_k +  \gamma_k +  O(\log n).
\end{equation}
\smallskip

We next analyze \emph{Case 2.} 
For (i),  as in Lemma~\ref{l:small}, we note that $x_k$ is a $(\hat{n}_k, \delta_k)$-rich owner w.r.t. $B = \{x \in \zo^n \mid C(x \mid b) \leq C(x_k \mid b) \}$.
We are now in the \emph{large regime case} and it follows that with probability $1-\delta_k$, $p_k$ has at most $(2/\delta^2_k) |B| 2^{\gamma_k} /2^{\hat{n}_k}$ neighbors in $B$, of which one is $x_k$.
Note that
\[
\frac{ (2/\delta^2_k)|B| 2^{\gamma_k}}{2^{\hat{n}_k}} \leq \frac{2^{C(x_k \mid b) + \gamma_k + \log(2/\delta^2_k) + 1}}{2^{\hat{n}_k}}  = 
 2^{C(x_k \mid b) -(n_k + \eta_k - \gamma_k - \log(2/\delta^2_k))}.
\]
So, $x_k$ can be constructed from $b, p_k, C(x_k \mid b)$ and the index of $x_k$ in an enumeration of $p_k$'s neighbors in $B$. This index is a string $b'$ of length at most $C(x_k \mid b) - (n_k +  \eta_k - \gamma_k - \log(2/\delta^2_k))$.
\smallskip

(ii) From part (i), with probability $1-\delta_k$,
\[
C(x_k \mid b) \leq C(p_k \mid b) + (C(x_k \mid b) - (n_k + \eta_k  - \gamma_k - \log(2/\delta^2_k)) + O(\log n).
\]
Therefore,
\begin{equation}
\label{e:pk}
C(p_k \mid b) \geq n_k  + \eta_k  - \gamma_k - \log(2/\delta^2_k) - O(\log n),
\end{equation}
which proves the second inequality in (ii).  Next, 
\[
\begin{split}
C(p_k, b) & \stackrel{(a)}{\geq}    C(b) + C(p_k\mid b) - O(\log n) \\
& \stackrel{(b)}{\geq} C(b) + n_k + \eta_k  - \gamma_k - \log(2/\delta^2_k) - O(\log n).
\end{split}
\]
Transition (a) follows by the chain rule and transition (b) uses~(\ref{e:pk}).
In the other direction, we have the inequality
\[
C(p_k,b) \leq C(b) + |p_k| + O(\log n) = C(b) + n_k  + \eta_k + \gamma_k + O(\log n).
\]
It follows that 
\begin{equation}
\label{e:eqcasetwo}
\big| C(p_k, b)  - (C(b) + n_k + \eta_k)\large| \leq \gamma_k +    \log(2/\delta^2_k) + O(\log n).
\end{equation}
Combining~(\ref{e:eqcaseone}) with ~(\ref{e:eqcasetwo}), the conclusion follows.
\end{myproof}

\begin{myproof} {\bf of Lemma~\ref{l:alg}.}
The computation is done iteratively in descending order for $k=\ell, \ell-1, \ldots, 1$.

At the first iteration $k=\ell$, there is nothing to compute because the values $C(x_V)$ are given, and thus the algorithm simply takes
$A(x_V) = C(x_V)$ for all $V \in {\cal P}([\ell])$. Note that $\gamma_{\ell+1} = 0$.

Suppose we have performed the iterations $\ell, \ell-1, \ldots, k$ and now we are at iteration $k-1$.

So we have already computed $A(x_{V'}, p_{[k+1..\ell]})$ for all non-empty $V' \subseteq [k]$ and with probability $1 - \hat{\delta}_{k+1}$,
\[
\big| C(x_{V'}, p_{[k+1..\ell]}) - A(x_{V'}, p_{[k+1..\ell]}) \big|  \leq 2 \gamma_{k+1}.
\]

Let us fix a non-empty $V \subseteq [k-1]$.  We will define $A(x_V, p_{[k+1..\ell]})$ (we do this below in equations~(\ref{e:axv}) and (\ref{e:axvl}) ). For this, we want to approximate $C(x_{V}, p_{k}, \ldots, p_\ell)$ because the plan is to use either Lemma~\ref{l:small}, (ii) or Lemma~\ref{l:large},(ii), with $b \leftarrow (x_V, p_{[k+1..\ell]})$. Which of the two lemmas is applicable depends on whether the complexity $C(x_k \mid b)$ is low  or high. Note that $C(x_k \mid b) = C(x_k, b) - C(b) \pm c \log n$ and at  the previous iteration we have computed the approximations $A(x_k, b)$  and $A(b)$ for $C(x_k,b)$ and respectively  $C(b)$. Therefore we distinguish two cases. 
\smallskip

{\bf Case 1 (low complexity case).} Suppose $A(x_{V \cup \{k\}}, p_{[k+1..\ell]}) -  A(x_{V}, p_{[k+1..\ell]}) \leq n_k + 8 \gamma_{k+1} + c \log n$.

 Note that with probability $1- \hat{\delta}_{k+1}$,  
 \[
 \begin{split}
 C(x_k \mid x_V, & p_{[k+1..\ell]})  \\
 & \stackrel{(a)}{\leq} C(x_{V}, x_k, p_{[k+1..\ell]}) - C(x_{V}, p_{[k+1..\ell]}) + c \log n \\ 
  & \stackrel{(b)}{\leq} A(x_{V}, x_k, p_{[k+1..\ell]}) - A(x_{V}, p_{[k+1..\ell]})+ 8 \gamma_{k+1} + c \log n\\
 & \stackrel{(c)}{\leq} n_k + 16 \gamma_{k+1} + 2c \log n \\
 & \stackrel{(d)}{\leq} n_k + \eta_k.
 \end{split}
 \]
Transition (a) follows by the chain rule, transition (b) uses the induction hypothesis, transition (c) uses the assumption that we are in Case 1, and transition (d) uses~(\ref{e:eta}) and~(\ref{e:gamma}).
 By Lemma~\ref{l:small} (ii) (with $b\leftarrow X_V, p_{[k+1..\ell]}$), with probability $1-\delta_k$,
 \begin{equation}
\label{e:cxv}
 \big| C(x_V, p_k, p_{[k+1..\ell]}) - C(x_V, x_k, p_{[k+1..\ell]}) \big| \leq \gamma_k + c \log n.
 \end{equation}
So, we define
\begin{equation}
\label{e:axv}
A(x_V, p_k, p_{[k+1..\ell]}) := A(x_{V}, x_k, p_{[k+1..\ell]}).
\end{equation}
Then, with probability $1- 2 \hat{\delta}_{k+1}- \delta_k =  1 - \hat{\delta_k}$,
\[
\begin{split}
\big| C(x_V, p_k, &p_{[k+1..\ell]}) - A(x_V,p_k, p_{[k+1..\ell]})  \big | \\
& \leq \big| C(x_V, p_k, p_{[k+1..\ell]}) - C(x_V, x_k, p_{[k+1..\ell]}) \big |   + 
\big | C(x_V, x_k, p_{[k+1..\ell]})  - A(x_V,p_k, p_{[k+1..\ell]}) \big | \\
& \stackrel{(a)}{=} \big| C(x_V, p_k, p_{[k+1..\ell]}) - C(x_V, x_k, p_{[k+1..\ell]}) \big |   + 
\big | C(x_V, x_k, p_{[k+1..\ell]})  - A(x_V, x_k, p_{[k+1..\ell]}) \big | \\
& \stackrel{(b)}{\leq} \gamma_k + c \log n  +4 \gamma_{k+1} \\
& \stackrel{(c)}{\leq} 4 \gamma_k.
\end{split}
\]
Transition (a) follows by~(\ref{e:axv}), transition (b) uses~(\ref{e:cxv}) and the induction hypothesis, and transition (c) uses~(\ref{e:gamma}).
\smallskip

{\bf Case 2 (high complexity case).} Suppose $A(x_{V \cup\{ k\}}, p_{[k+1..\ell]}) -  A(x_{V}, p_{[k+1..\ell]}) >  n_k + 8 \gamma_{k+1} + c \log n$.

 This time, with probability $1- 2 \hat{\delta}_{k+1}$,  
 \[
 \begin{split}
 C(x_k \mid x_V, & p_{[k+1 ..\ell]})  \\
 &\stackrel{(a)}{\geq} C(x_{V \cup \{k\}}, p_{[k+1..\ell]}) - C(x_{V}, p_{[k+1..\ell]})  - c \log n \\ 
 &\stackrel{(b)}{\geq} A(x_{V \cup \{k\}}, p_{[k+1..\ell]}) - A(x_{V}, p_{[k+1..\ell]}) - 8\gamma_{k+1} - c \log n\\
 &\stackrel{(c)}{\geq}  n_k.
 \end{split}
 \]
Transition (a) follows by the chain rule, transition (b) uses the induction hypothesis, and transition (c) uses the assumption that we are in Case 2.
 By Lemma~\ref{l:large} (ii), with probability $1-\delta_k$,
 \begin{equation}
\label{e:cxvl}
 \big| C(x_V, p_k, p_{[k+1..\ell]}) - (C(x_V,  p_{[k+1..\ell]}) + n_k + \eta_k ) \big | \leq  \gamma_k + \eta_k + \log (2/\delta^2_k)+  c \log n.
 \end{equation}
So, we define
\begin{equation}
\label{e:axvl}
A(x_V, p_k, p_{[k+1..\ell]}) := A(x_{V}, p_{[k+1..\ell]})  + n_k + \eta_k.
\end{equation}
Then, with probability $1- 2\hat{\delta}_{k+1} - \delta_k \geq 1 - \hat{\delta}_k$, 
\[
\begin{split}
\big| C(x_V, p_k, & p_{[k+1..\ell]} - A(x_V,p_k, p_{[k+1..\ell]})  \big | \\
&\leq \big| C(x_V, p_k, p_{[k+1..\ell]}) - (C(x_V, p_{[k+1..\ell]}) +n_k + \eta_k) \big |   +  \\
& \quad\quad + \big | (C(x_V, p_{[k+1..\ell]}) + n_k + \eta_k)   - A(x_V, p_k, p_{[k+1..\ell]}) \big | \\
&\stackrel{(a)}{\leq} \big| C(x_V, p_k, p_{[k+1..\ell]}) - (C(x_V, p_{[k+1..\ell]}) +n_k + \eta_k) \big |   + \\
& \quad\quad + \big | (C(x_V, p_{[k+1..\ell]}) + n_k + \eta_k)   - (A(x_V,  p_{[k+1..\ell]})+n_k +\eta_k) \big | \\
&\stackrel{(b)}{\leq} \gamma_k + \eta_k + \log (2/\delta^2_k) + c \log n + 4 \gamma_{k+1} \\
&\stackrel{(c)} = \gamma_k + 2\gamma_{k}  + 2 \log (2/\delta^2_k) + c \log n + 4 \gamma_{k+1} \\
& \stackrel{(d)}{\leq} 4 \gamma_k.
\end{split}
\]
Transition (a) follows by~(\ref{e:axvl}), transition (b) uses~(\ref{e:cxvl}) and the induction hypothesis, transition (c) uses~(\ref{e:eta}), and transition (d) uses~(\ref{e:gamma}).
\end{myproof}

\begin{myproof} {\bf of Lemma~\ref{l:bound}.}
We do backward induction on $k$.
The statement is true for $k = \ell$, by hypothesis.
Suppose we have proven the statement for $k+1$. We prove it for $k$.
Let $V \subseteq [k]$.
\smallskip

{\bf Case 1 (low complexity case).} Suppose $C(x_{k+1} \mid x_{[k]-V}, p_{[k+2 ..\ell]}) \leq n_{k+1} + \eta_{k+1}$.

We apply Lemma~\ref{l:small} for $k+1$ and $b := x_{[k]-V}, p_{[k+2..\ell]}$. We obtain that, with probability $1-\delta_{k+1}$,  $x_{k+1}$ can be constructed from $p_{k+1}, b$ and $C(x_{k+1} \mid b)$.

Next,
\[
\begin{split}
C(x_V \mid x_{[k]-V}, p_{[k+1..\ell]})  &\stackrel{(a)}{\leq} C(x_V \mid x_{[k]-V}, x_{k+1}, p_{[k+2..\ell]}) + c \log n  \\
&\stackrel{}{=} C(x_V \mid x_{[k+1]-V}, p_{[k+2..\ell]}) + c \log n \\
& \stackrel{(b)}{\leq} \sum_{j \in V} n_j + O(\ell-k-1) \cdot \log n + c \log n \\
& = \sum_{j \in V} n_j + O(\ell-k) \cdot \log n.
\end{split}
\]
Transition (a) holds by the above argument with probability $1-\delta_k$, transition (b) holds by the induction hypothesis with probability $1-\hat{\delta}_{k+1}$. Thus the entire chain of inequalities holds with 
probability $1-\hat{\delta}_{k+1} - \delta_{k+1} \geq 1 - \hat{\delta_k}$.
\smallskip

{\bf Case 2 (high complexity case).} Suppose $C(x_{k+1} \mid x_{[k]-V}, p_{[k+2..\ell]}) >  n_{k+1} + \eta_{k+1}$.

Then, Lemma~\ref{l:large}, used for $k+1$ and $b:=  x_{[k]-V}, p_{[k+2..\ell]}$, implies that with probability $1-\delta_{k+1}$,
\begin{equation}
\label{e:aa}
\begin{split}
C(p_{k+1} \mid x_{[k]-V}, p_{[k+2 ..\ell]}) & \geq n_{k+1} + \eta_{k+1} -\gamma_{k+1} - \log (2/\delta^2_{k+1})  - c \log n. 
\end{split}
\end{equation}
Next,
\[
\begin{split}
C(x_V & \mid x_{[k]-V}, p_{[k+1..\ell]})  \\
& \stackrel{(a)}{\leq} C(x_V, p_{k+1} \mid x_{[k]-V}, p_{[k+2..\ell]}) - C(p_{k+1} \mid x_{[k]-V}, p_{[k+2..\ell]})+ c \log n \\ 
& \stackrel{(b)}{\leq}  C(x_V, p_{k+1} \mid x_{[k]-V},p_{[k+2..\ell]}) - n_{k+1} - \eta_{k+1} + \gamma_{k+1} + \log (2/\delta^2_{k+1})+ 2c \log n \\ 
& \stackrel{(c)}{\leq} C(x_V, x_{k+1} \mid x_{[k]-V}, p_{[k+2..\ell]}) - n_{k+1}  - \eta_{k+1} + \gamma_{k+1} + \log (2/\delta^2_{k+1}) + \gamma_{k+1} + 2c \log n \\ 
&\stackrel{ (d)}{\leq}C(x_{V \cup\{k+1\}} \mid x_{[k+1] - V \cup\{k+1\}}, p_{[k+2..\ell]}) - n_{k+1} + 2c \log n \\ 
&\stackrel{(e)}{\leq} \big(\sum_{j \in V \cup \{k+1\}} n_j) + O(\ell-k-1) \cdot \log n - n_{k+1} + 2c \log n \\ 
&\stackrel{ }{=} \sum_{j \in V} n_j + O(\ell-k) \cdot \log n.
\end{split}
\]
Transition (a) follows by the chain rule, transition (b) follows from inequality~(\ref{e:aa}) and holds with probability $1-\delta_{k+1}$, transition (c) follows from the fact that $p_{k+1}$ can be obtained from $x_{k+1}$ and the 
index of the edge that connects $x_{k+1}$ and $p_{k+1}$ and this index needs $\gamma_{k+1}$ bits, transition (d) holds due to~(\ref{e:eta}).
Inequality (e) holds with probability $1- \hat{\delta}_{k+1}$ by the induction hypothesis for $k+1$. Taking into account transitions (a) and (e), the entire chain of inequalities holds with 
probability $1-\hat{\delta}_{k+1} - \delta_{k+1} \geq 1 - \hat{\delta_k}$.
\end{myproof}

\section{Construction of graphs with the rich owner property}
\label{s:graph}
In this section we prove Theorem~\ref{t:richownergraph}.
The construction relies on the  randomness extractor of  Raz, Reingold, and Vadhan~\cite{rareva:c:extractor}. We recall that a $(k, \epsilon)$ extractor is a function $E: \zo^n \times \zo^d \rightarrow  \zo^m$ such that for any distribution $X$ on $\zo^n$ with min-entropy $H_\infty(X) \geq  k$, $E(X,U_d)$ is $\epsilon$-close to $U_m$, where $U_d$ ($U_m$) is the uniform distribution on $\zo^d$ (respectively, $\zo^m$), i.e., for every $A \subseteq \zo^m$,
\begin{equation}
 \label{eq:extractorDef}
\bigg| \prob[E(X, U_d) \in A] - \frac{|A|}{M} \bigg| < \epsilon.
\end{equation}

\begin{theorem}[Th.1, (2) in ~\cite{rareva:c:extractor}]
\label{t:rrv} Let $k(n) \leq n$ and $1/\epsilon(n)$ be functions mapping natural numbers to natural numbers and computable in polynomial time.
There exists a family of functions $E_n: \zo^n \times \zo^{d(n)}  \rightarrow \zo^{k(n)}$ computable uniformly in polynomial time such that
\begin{itemize}
\item[(1)] For every $k' \leq k(n)$, the prefix $k'$ of $E$ (i.e., the function obtained by computing $E$ and retaining only the first $k'$ bits of the output) is a $(k',\epsilon)$ extractor,
\item[(2)] $d(n) = O(\log^2(n/\epsilon(n)) \log n)$.
\end{itemize}
\end{theorem}
Next we convert the extractor from Theorem~\ref{t:rrv} into a graph with the rich owner property. The method follows closely~\cite{bau-zim:c:linlist}.  We first establish several  lemmas. 


Let $E: \zo^n \times \zo^d \mapping \zo^m$ be a $(k,\epsilon)$ extractor, and let $G_E = (L =\zo^n, R=\zo^m, E_G)$ be the corresponding bipartite graph, \ie, there is an edge $(x,z) \in E_G$ iff there exists $y$ such that $E(x,y)=z$.
 Let $B \subseteq L$.  The $B$-degree of a node $y$  (denoted ${\rm deg}_B(y)$) is the number of $y$'s neighbors that are in $B$. Let $D = 2^d$. 

 A vertex $y \in R$ is $t$-heavy for $B$ if ${\rm deg}_B(y) \geq  t \cdot \frac{|B|\cdot D}{|R|}$ (otherwise $y$ is $t$-light for $B$).

 Let $A = |\{ y \in R \mid  y \mbox{ is $t$-heavy for $B$}\}$.

A vertex $x \in B$ is $\delta$-bad for $B$ if ${\rm deg_A(x)}/{\rm deg(x)} \geq \delta$ (\ie, more than a $\delta$ fraction of edges outgoing from $x$ land in nodes that are $t$-heavy for $B$).

\begin{lemma}
\label{l:heavy}
For every bipartite graph $G$, for every $B \subseteq L$, for every $t > 0$,
$|A| \leq \frac{1}{t}\big |R \big |$. 

\end{lemma}
\begin{myproof}
 The number of edges between $B$ and $A$ is at least $|A| \cdot t \cdot \frac{ |B|\cdot D}{|R|}$. On the other hand, the total number of edges between $B$ and $R$ is $|B| \cdot D$. Thus, $|A| \cdot t \cdot  \frac{|B|\cdot D}{|R|} \leq |B| \cdot D$, from which the conclusion follows.
\end{myproof}
\smallskip

Let $\deb$ be the set of vertices in $B$ which are $\delta$-bad for $B$. 

\begin{lemma}
\label{l:bad} If $|B| \geq 2^k$, then 
$\frac{|\deb|}{|B|} \leq  \frac{1}{\delta}\bigg (\frac{1}{t} + \epsilon \bigg)$.

\end{lemma}
\begin{myprooftwo}
Let $X$ be the distribution which is flat on $B$ (\ie, it assigns equal probability mass to elements in $B$, and $0$ probability mass to every element which is not in $B$). Then, $H_\infty(X) \geq k$.
Let $\mu_E$ be the  distribution induced by the extractor $E$ on $R$ when $x$ is chosen according to distribution $X$ and $y$ is chosen uniformly at random in $\zo^d$.
 Formally, for $Z \subseteq R$, 
\[
\mu_E(Z) = \frac{|\{(x,y) \mid x \in B, y \in \zo^d, E(x,y) \in Z\}|}{|B|\cdot D}.
\]
Since $E$ is $(k, \epsilon)$-extractor, $\mu_E(A) \leq \frac{|A|}{|R|} + \epsilon \leq \frac{(1/t)|R|}{|R|} + \epsilon = \frac{1}{t} + \epsilon$.

On the other hand, $\mu_E(A) \geq \frac{|\deb| \cdot \delta D}{|B| \cdot D} = \frac{|\deb|}{|B|} \cdot \delta$.

So,
$\frac{|\deb|}{|B|} \leq \frac{1}{\delta} \cdot \mu_E(A) \leq \frac{1}{\delta}\bigg (\frac{1}{t} + \epsilon \bigg). $\hfill \qed

\end{myprooftwo}
\if01
\begin{lemma}[~\cite{bau-zim:c:linlist}]
\label{l:mod}
Let $x_1, x_2 \ldots, x_s$ be distinct $n$-bit strings, which we view in some canonical way as integers $< 2^{n+1}$.

Let $p_i$ be the $i$-th prime number and let $L = \{p_1, \ldots, p_t\}$, where $t = (1/\delta) \cdot s \cdot n$.

For every $i \leq s$, for $(1-\delta)$ fraction of $p$ in $L$, the value of $x_i \hspace{-0.2cm}\mod p$ is unique in the sequence 
$(x_1\hspace{-0.2cm} \mod p, x_2 \hspace{-0.2cm}\mod p, \ldots,  x_s\hspace{-0.2cm} \mod p)$.
\end{lemma}
\fi
Now we describe the transformation of an extractor graph into a graph with the rich owner property.  We use again the hashing technique provided by Lemma~\ref{l:mod}.
\medskip

\begin{mdframed}
Let $s$ be a positive integer and let $\delta >0$. The following  algorithm transforms  $G_1 = (L = \zo^n, R_1 = \zo^m, E_1)$, a bipartite graph  into another bipartite graph $G$ as follows.

Let $\ell = (1/\delta) \cdot s \cdot n$ and let $q_1, q_2, \ldots q_\ell$ be the first $\ell$ prime numbers.
We construct the bipartite graph 
\[G = (L =\zo^n, R = \{q_1, \ldots, q_\ell\} \times \{0,1, \ldots, q_\ell-1\} \times R_1, E),
\]
 by adding for each $(x,z)$ in $E_1$ the edges 
\[
(x, (q_1, x \bmod q_1, z)),  (x, (q_2, x \bmod q_2, z)), \ldots, (x, (q_\ell,  x \bmod q_\ell, z))
\]
 in $E$ (one can think that each edge  $(x, z) \in G_1$ is split into $\ell$  edges in $G$).
\end{mdframed}

\begin{lemma}
\label{l:two}
Let  $G_1 = (L_1 = \zo^n, R_1 = \zo^k, E_1 \subseteq L_1 \times R_1)$  be the bipartite graph with left degree $D_1 = 2^{d_1}$  corresponding to the function $E_n$  from Theorem~\ref{t:rrv} with parameters $k=k(n)$ and $\epsilon = \epsilon(n)$.  Let $\delta = (2\epsilon)^{1/2}$ and let  $G = (L,R, E \subseteq L \times R)$ be  constructed from $G_1$ as above with $s= (2/\delta^2) \cdot 2^{d_1}$. Then:

\begin{itemize}
\item[(1)] $G$ has the $(k, 2\delta)$- rich owner property.
\item[(2)] $L = \zo^n$.
\item[(3)]  $R$ can be taken to be $\zo^{3 \log \ell} \times \zo^k$, where $\ell = (1/\delta)\cdot s \cdot n$.
\item[(4)] The left degree of $G$ is bounded by $2^d \cdot \ell$. 
\item[(5)] If $G_1$ is explicit, then $G$ is explicit.
\end{itemize}
\end{lemma}
\begin{myproof}
We analyze first the \emph{small regime case}. Let $B \subseteq L$ be  a subset of size $2^{k'} \leq 2^k$ (to simplify the notation we assume that the size of $B$ is a power of two).
We consider the graph $G_1'$, the $k'$-prefix of $G_1$, which means that $G_1'$ is obtained from $G_1$ by reducing the labels of the right nodes from their initial $k$-bit value to the prefix of length $k'$.  By Theorem~\ref{t:rrv}, $G_1'$ is a $(k', \epsilon)$ extractor.
  By Lemma~\ref{l:bad} (in which we take $\epsilon = \delta^2/2$ and $t=2/\delta^2$), there is a ``bad" set
$\deb \subseteq B$ of size $|\deb| \leq \delta |B|$, such that for all the ``good" nodes
$x \in B - \deb$,  in $G_1'$, it holds that at least $(1-\delta)$ fraction of edges outgoing from $x$ land in right nodes that are $t$-light for $B$, \ie, land in right nodes that have $B$-degree in $G_1'$  at most   $(2/\delta^2) \cdot  |B| \cdot D_1 / |R_1'| 
= (2/\delta^2) \cdot (2^{k'+d_1-k'}) = s$. The $B$-degree of a node in $G_1$ can be at most  the $B$-degree of its prefix in $G_1'$, and therefore the above holds in $G_1$ as well.

Let us fix a ``good" node $x \in B -  \deb$. Suppose the multiset of $x$'s neighbors in $G_1$ is
$\{z_1, z_2, \ldots, z_D\}$.  We write the neighbors of $x$ in $G$ in the following tabular form:
\[
\begin{array}{c c c c}
( q_1, x \bmod q_1, z_1) &  (q_2,  x \bmod q_2, z_1) & \ldots & ( q_\ell,  x \bmod q_\ell, z_1) \\

(q_1,  x \bmod q_1, z_2) &  (q_2, x \bmod q_2, z_2) & \ldots & (q_\ell, x \bmod q_\ell, z_2) \\

\vdots & & & \\

(q_1, x \bmod q_1, z_D) &  (q_2, x \bmod q_2, z_D) & \ldots & (q_\ell, x \bmod q_\ell, z_D)

\end{array}
\]
In at least a fraction of $(1-\delta)$ rows, the corresponding $z_i$  has $deg_B(z_i) \leq s$ in $G_1$, so each node in such a row is shared by at most $s$ elements of $B$, say $x, x_2, \ldots, x_s$.  In each such row, if we look at the components $q_i, x \bmod q_i$ and 
take into account Lemma~\ref{l:mod}, we conclude that at least a fraction $(1-\delta)$ of the elements in the row have a unique neighbor in $B$ (in $G$). Thus, overall, at least a fraction of $(1-\delta)^2 > (1-2\delta)$ of the neighbors of $x$ are unique.  
Since this holds for every $x \in B - \deb$ and $|B-\deb| = |B| - |\deb| \geq (1-\delta)|B| > (1-2\delta)|B|$, we are done.

Next, we analyze the \emph{large regime case}. Let $B \subseteq L$ be  a subset of size $2^{k'} >  2^k$.   By Lemma~\ref{l:bad} (in which again we take $\epsilon = \delta^2/2$ and $t=2/\delta^2$), there is a ``bad" set
$\deb \subseteq B$ of size $|\deb| \leq \delta |B|$, such that for all the ``good" nodes
$x \in B - \deb$,  in $G_1$, it holds that at least $(1-\delta)$ fraction of edges outgoing from $x$ land in right nodes that are $t$-light for $B$, \ie, they are shared with at most $(2/\delta^2) \cdot  |B| \cdot D_1 / |R| 
= (2/\delta^2) \cdot |B| \cdot D_1 / 2^k$ other nodes from $B$.  The edge splitting operation can only reduce congestion, and the left degree increases from $D_1$ to $D = D_1 \cdot \ell$. So,  in $G$ it holds that 
 all the ``good" nodes
$x \in B - \deb$,  have at least a $(1-\delta)$ fraction of edges outgoing from $x$ that land in right nodes that  are shared with at most $(2/\delta^2) \cdot  |B| \cdot D / 2^k$ 
 other nodes from $B$.  Since $|B-\deb| = |B| - |\deb| \geq (1-\delta)|B| > (1-2\delta)|B|$, we are done.

The parameters of $G$ follow from its   construction taking into account that $q_\ell \leq \ell \log \ell$ and that the $\ell$'s prime number can be found in time polynomial in $\ell$.
\end{myproof}

The proof of Theorem~\ref{t:richownergraph} follows immediately from Lemma~\ref{l:two}.

\section{Acknowledgments} The author is grateful to Nikolay Vereshchagin and Alexander Shen for useful discussions. Sasha Shen suggested the idea that led to the elimination in the main result of the requirement that the the inputs' complexity profile must be given to the decompression procedure. The author thanks his father, Rudy Zimand, 
for helping him with the Russian language  in~\cite{rom:j:slepwolf} (an English version is now available~\cite{rom:t:slepwolf}).

\bibliography{$HOME/Documents/theory}

\bibliographystyle{alpha}

\end{document}